# Spin structure and dynamics of the topological semimetal Co$_3$Sn$_{2-x}$In$_x$S$_2$

Kelly J. Neubauer[1], Feng Ye[2], Yue Shi[3], Paul Malinowski[3], Bin Gao[1], Keith M. Taddei[2], Philippe Bourges[4], Alexandre Ivanov[5], Jiun-Haw Chu[3,#], and Pengcheng Dai[1,*]

[1]Department of Physics and Astronomy, Rice University, Houston, Texas 77005, USA

[2]Neutron Scattering Division, Oak Ridge National Laboratory, Oak Ridge, Tennessee 37831, USA

[3]Department of Physics, University of Washington, Seattle, Washington 98195, USA

[4]Laboratoire Léon Brillouin, CEA-CNRS, Université Paris-Saclay, CEA Saclay, 91191 Gif-sur-Yvette, France

[5]Institut Laue-Langevin, 71 avenue des Martyrs CS 20156, 38042 Grenoble Cedex 9, France

[*]email: pdai@rice.edu; [#]email: jhchu@uw.edu

**Abstract**

The anomalous Hall effect (AHE), typically observed in ferromagnetic (FM) metals with broken time-reversal symmetry, depends on electronic and magnetic properties. In Co$_3$Sn$_{2-x}$In$_x$S$_2$, a giant AHE has been attributed to Berry curvature associated with the FM Weyl semimetal phase, yet recent studies report complicated magnetism. We use neutron scattering to determine the spin dynamics and structures as a function of $x$ and provide a microscopic understanding of the AHE and magnetism interplay. Spin gap and stiffness indicate a contribution from Weyl fermions consistent with the AHE. The magnetic structure evolves from $c$-axis ferromagnetism at $x = 0$ to a canted antiferromagnetic (AFM) structure with reduced $c$-axis moment and in-plane AFM order at $x = 0.12$ and further reduced $c$-axis FM moment at $x = 0.3$. Since noncollinear spins can induce non-zero Berry curvature in real space acting as a fictitious magnetic field, our results revealed another AHE contribution, establishing the impact of magnetism on transport.

**Introduction**

The concept of topology has been predicted and experimentally identified amongst materials with little or no electron correlations and is arguably a success in the pursuit of materials by design [1, 2]. In contrast, topology in strongly correlated materials is much less explored due to a lack of identified material platforms and the difficulty in developing theories that incorporate topology and electron correlations. The role of magnetism in topologically protected states has implications for both our fundamental understanding of properties and the technological applications of quantum materials such as dissipationless spintronics [3, 4]. The magnetic semimetal $Co_3Sn_2S_2$ is particularly interesting because the interplay between magnetism and topology leads to giant anomalous Hall effect (AHE), where charge carriers acquire a velocity component orthogonal to an applied electric field without an external magnetic field, with a small ordered moment [5, 6]. Remarkably, the size of AHE in $Co_3Sn_2S_2$ is among the largest compared to most previously reported AHE materials [5]. The magnetic Co ions in $Co_3Sn_2S_2$ form a two-dimensional (2D) kagome lattice, composed of corner sharing triangles and hexagons, separated by nonmagnetic S and Sn layers with another Sn intercalated between the Co-S layer [Fig. 1a]. Depending on the nature of the magnetic order, the observed AHE may have different microscopic interpretations [7-12]. If $Co_3Sn_2S_2$ is a simple half-metallic ferromagnet below $T_C = 177$ K with Co moment aligned along the *c*-axis as predicted by *ab initio* calculations, the spin-orbit coupling (SOC) would induce a splitting of the electronic bands and open up gaps with the anti-crossing nodal line connected to the Weyl points [5, 6, 13-17]. In this case, the intrinsic AHE in Co$_3$Sn$_2$S$_2$ arises from Berry curvature of the pairs of Weyl points characterized by the opposite chiralities acting as the monopole and anti-monopole of the emergent magnetic field in momentum space. However, if the magnetic structure has an additional in-plane $120°$ antiferromagnetic (AFM) order component with one basis vector (termed $\psi_1$) below ~90 K as suggested by muon-spin rotation (μSR) measurements [the red arrows in Fig. 1b] [11], the numbers of Weyl nodes and their locations in reciprocal space will change considerably, thus giving rise to different emergent magnetic field, Berry curvature, and intrinsic AHE (called topological Hall effect or THE). Although recent polarized neutron diffraction experiments on Co$_3$Sn$_2$S$_2$ ruled out the $\psi_1$ AFM order proposed in Ref. [11], a different in-plane $120°$ AFM order with two basis vectors as shown in blue arrows of Fig. 1b ($\psi_2$) is still possible though not confirmed [12].

One way to resolve the magnetic structure of Co$_3$Sn$_2$S$_2$ is to carry out systematic neutron scattering experiments as a function of chemical doping and compare the outcome with AHE measurements. By partially substituting Sn with In to form Co$_3$Sn$_{2-x}$In$_x$S$_2$, the system changes from an FM Weyl semimetal to a nonmagnetic insulator [18]. While the low-temperature magnetization of the system decreases monotonically with increasing In-doping, the AHE enhances significantly near $x = 0.15$ and decreases until vanishing at $x > 0.8$ [19]. This non-monotonic doping dependence of the intrinsic anomalous Hall conductivity in Co$_3$Sn$_{2-x}$In$_x$S$_2$ has been attributed to doped induced changes in Berry curvature near the Fermi level [19]. On the other hand, much is unknown about the evolution of the magnetic structure of this system. For example, magnetic, transport, and optical measurements on $Co_3Sn_2S_2$ reveal a clear anomaly around 120 K well below $T_C = 177$ K [11, 15, 16, 20]. Although previous neutron scattering experiments could not rule out the $\psi_2$ AFM order, a sudden reduction in FM domain size is believed to be responsible for the observed properties below 120 K [12].

In this paper, we report neutron scattering experiments to determine the evolution of spin dynamics and magnetic structures in $Co_3Sn_{2-x}In_xS_2$ as a function of $x$ that provide a microscopic understanding of the doping and temperature dependent AHE. We find that the magnon gap due to SOC and in-plane spin stiffness are exhibit a non-monotonous temperature dependence that follows the enhanced AHE as a function of temperature and In-doping. Additionally, we determine the evolution of the magnetic structure as a function of temperature and In-doping $x$. With increasing $x$, the magnetic structure changes from a simple ferromagnet with moment along the $c$-axis at $x = 0$ to a canted AFM structure with reduced $c$-axis FM moment and an in-plane $120°$ AFM order with two basis vectors at $x = 0.12$. The AFM moment increases coupled with decreasing FM moment with increasing doping until magnetism disappears at $x = 1$. Since noncollinear spin texture can induce non-zero Berry curvature acting as fictitious magnetic field for the conduction electrons [21-23], our results suggest that there is an additional contribution to the AHE, thus establishing the basis for a comprehensive understanding of AHE in these materials.

**Results**

**Preliminary characterization** Large single crystals of Co$_3$Sn$_{2-x}$In$_x$S$_2$ with $x = 0, 0.12, 0.3$ were grown using a flux method [24]. Their compositions were confirmed using powder X-ray diffraction and Inductively Coupled Plasma Optical Emission Spectroscopy at Rice University (see Methods Section for details). Neutron diffraction measurements were performed on single crystals with $x = 0.12, 0.3$ using the elastic diffuse scattering spectrometer CORELLI at the Spallation Neutron Source (SNS) and the WAND$^2$ diffractometer of the High Flux Isotope Reactor both at Oak Ridge National Laboratory (ORNL) [25, 26]. Co$_3$Sn$_{2-x}$In$_x$S$_2$ crystallizes in a rhombohedral structure (space group: $R\bar{3}m$) with Co atoms that form kagome layers in the *ab*-plane [Fig. 1a]. The corresponding Brillouin zone is shown in Figs. 1c,d. We define the momentum transfer $\boldsymbol{Q}$ in 3D reciprocal space in Å$^{-1}$ as $\boldsymbol{Q} = H\boldsymbol{a}^* + K\boldsymbol{b}^* + L\boldsymbol{c}^*$, where *H*, *K*, and *L* are Miller indices and $\boldsymbol{a}^* = 2\pi(\boldsymbol{b} \times \boldsymbol{c})/[\boldsymbol{a} \cdot (\boldsymbol{b} \times \boldsymbol{c})]$, $\boldsymbol{b}^* = 2\pi(\boldsymbol{c} \times \boldsymbol{a})/[\boldsymbol{a} \cdot (\boldsymbol{b} \times \boldsymbol{c})]$, $\boldsymbol{c}^* = 2\pi(\boldsymbol{a} \times \boldsymbol{b})/[\boldsymbol{a} \cdot (\boldsymbol{b} \times \boldsymbol{c})]$ with $\boldsymbol{a} = a\hat{\boldsymbol{x}}$, $\boldsymbol{b} = a(\cos 120\,\hat{\boldsymbol{x}} + \sin 120\,\hat{\boldsymbol{y}})$, $\boldsymbol{c} = c\hat{\boldsymbol{z}}$ ($a = b \sim 5.36$, $c \sim 13.14$ Å) [Figs. 1a-d].

Figures 1e and 1f show the temperature dependence of the *c*-axis and in-plane magnetic susceptibilities, respectively, for $x = 0.12, 0.3$. The applied field is 0.1 T, slightly lower than the 0.5 T field used in Ref. [19]. While the overall In-doping dependent data is consistent with Ref. [19], we find a clear reduction in the *c*-axis field magnetic susceptibility below 50 K for $x = 0.12$, suggesting the presence of a AFM component. The signature of FM domain size reduction, seen most clearly as a kink in the in-plane field susceptibility data below ~120 K for $x = 0$ [12], is confirmed for $x = 0.12, 0.3$ single crystals [Fig. 1f].

Figures 1g and 1h summarize the In-doped dependence of the AHE and its comparison with the magnetic structures of the system obtained from our transport and neutron scattering experiments and Ref. [19]. Assuming $Co_3Sn_2S_2$ is a simple ferromagnet with *c*-axis aligned moment [Fig. 1i], increasing In-doping to $x = 0.12$ changes the magnetic structure to canted antiferromagnet with *c*-axis moment and in-plane $\psi_2$ AFM order [Fig. 1j]. Upon further increasing In-doping to $x = 0.3$, the FM component along the *c*-axis decreases and the in-plane AFM moment increases while maintaining the same AFM structure [Fig. 1j].

**Neutron diffraction** For each doping level on CORELLI, a single crystal of 10-30 mg was mounted on an aluminum pin and aligned in the scattering plane. The crystal was rotated through 360 degrees to cover a large area of reciprocal space. For $x = 0.12$, full maps were collected at 6 [Fig. 2a], 60, and 165 K [Fig. 2b]. For $x = 0.3$, full maps of neutron intensities were collected at 6 and 160 K. The data were reduced using MANTID [27] to extract nuclear and magnetic Bragg peak intensities. The nuclear and magnetic structure refinements were then carried out using Jana2006 software [28]. Figures 2a,b show reciprocal space maps in the $[H, 0, L]$ scattering plane for the $x = 0.12$ at $T = 6$ and 165 K obtained on CORELLI. From susceptibility measurements shown in Figs. 1e and 1f, we know that FM component of susceptibility decreases below around 50 K, in addition to changes in FM domain around 120 K. Figure 2c shows the temperature dependence of the (1,0,1) magnetic Bragg peak for the $x = 0.12$ sample. Below $T_C = 165$ K, the intensity of the (1,0,1) peak increases with decreasing temperature showing no anomaly around 120 K and reaches a maximum near 50 K before sharply decreasing upon further decreasing temperature. This decrease indicates a deviation from normal FM behavior and an increase in the in-plane AFM component in the magnetic structure.

From the full reciprocal space map at 165 K ($T > T_C$), the crystal lattice structure was refined and used as the input for the magnetic refinement. We consider the symmetry allowed magnetic structures compatible with the magnetic Co atom positions and $\mathbf{k} = 0$ propagation vector as there are no new superlattice peaks appearing below . The possible basis vectors are detailed in the Methods Section. At $T = 60$ K ($> T_N$), the magnetic structure can be well refined assuming a simple $c$-axis aligned FM (the $\psi_3$ structure) [Fig. 2e]. On cooling to temperatures below 50 K, when the system becomes AFM, we can refine the magnetic structure assuming coexistence of a $c$-axis aligned FM moment with the $\psi_1$ or $\psi_2$ in-plane AFM structure. Both configurations have 120-degree, in-plane AFM moments with $\psi_1$ moments along ($a,a,0$), (-$a,0,0$), and ($0,-a,0$) and while $\psi_2$ has moments along ($a,-a,0$), ($a,2a,0$), and (-$2a,-a,0$). Consistent with Ref. [12], we find that the $\psi_1 + \psi_3$ magnetic structure cannot best explain the data. On the other hand, the $\psi_2 + \psi_3$ magnetic structure with $\varphi = 25°$ gives a satisfactory refinement to the data, forming a canted AFM or umbrella-like structure with components along both the $c$-axis and in-plane, $\psi_2$ AFM order [Figs. 2d,g].

For $x = 0.3$, the nuclear structure at 160 K was refined and used as an input for the magnetic structure refinement at 6 K. The refinement calculated intensities fit well with the measured intensities as shown in Fig. 2i. Our refinements indicate that the best fit to the data is again the $\psi_2 + \psi_3$ magnetic structure but with $\varphi = 65°$ at 6 K [Fig. 2i]. Therefore, we conclude that as In-doping is increased from 0 to 0.3, the low temperature magnetic structure evolves from a simple FM phase in the parent compound to a canted FM with components along both the $c$-axis and in-plane AFM order, where the AFM moment size increases with In-doping at the expense of the FM moment. In-doping introduces holes to the system and slightly increases the separation of kagome layers which may explain why this system disfavors FM arrangement with increased doping [11, 29].

**Magnetization and transport measurements** From the magnetic susceptibility [Fig. 1e] and order parameter measurements [Fig. 2c], we see that the non-collinear canted spin structure only appears below about 50 K in the $x = 0.12$ sample. Since the $\psi_2 + \psi_3$ magnetic structure is noncollinear and has non-zero spin chirality [$\chi = \mathbf{S}_i \cdot (\mathbf{S}_j \times \mathbf{S}_k) \neq 0$, where $\mathbf{S}_i$, $\mathbf{S}_j$, $\mathbf{S}_k$ are the three nearest spins], it can induce non-zero Berry curvature acting as fictitious magnetic field for the conduction electrons to give rise to the THE [30-33]. Recent neutron diffraction work on YMn$_6$Sn$_6$ showed field-induced non-collinear spin can indeed induce THE [34-36]. To test if this is also the case for Co$_3$Sn$_{2-x}$In$_x$S$_2$, we carried out *c*-axis magnetic field dependence of the Hall resistivity for $x = 0.12$ and 0.3 samples [Figs. 3a,b] [19]. The magnetization measurements for these samples are shown in Fig. 1e with anomalous low temperature behavior in the $x = 0.12$ sample. We clearly see an anomalous low temperature behavior of anomalous Hall conductivity (AHC) $\sigma_{xy}(T)$ for the $x = 0.12$ sample in Fig. 3c. We compare the AHC and magnetization of the $x = 0.12$ sample in Fig. 3d and observe a slight decrease followed by an increase in AHC below 50 K despite a monotonic decrease in magnetization. Therefore, the enhanced AHC at low temperature must be due to an additional mechanism such as non-zero spin chirality. We attribute this low temperature effect to the THE whose presence is in addition to AHE due to net magnetization which are proportional to $a\rho_{xx}M + b\rho_{xx}^2 M$. In Fig. 3e, we show the anomalous Hall resistivity and in Fig. 3f we fit the results to the AHE contribution due to the net magnetization as discussed above. In the $x = 0.12$ sample, we see an extra contribution that we attribute to THE due to the spin-chirality of the refined magnetic structure as indicated by the shaded region where an approximately 10% deviation is observed. This 10% deviation corresponds to a 3.5 µΩ cm topological resistivity, which is comparable to the size of THE in other well know non-colinear antiferromagnets such as Mn$_3$Sn [37] and Nd$_2$Mo$_2$O$_7$ [38].

**Inelastic neutron scattering** Having determined the magnetic structure of Co$_3$Sn$_{2-x}$In$_x$S$_2$, we now consider the effect of spin dynamics to the AHE [39-41]. In previous work on Weyl semimetal candidate SrRuO$_3$, the observed non-monotonic temperature dependence of the AHC $\sigma_{xy}(T)$, induced by Berry curvatures near Weyl points, is associated with temperature dependence of the spin gap $E_g(T)$ in FM spin waves via:

$$E_g(T) = \frac{a_g M(T)/M_0}{1 + b(\frac{M(T)}{M_0})(\frac{\sigma_{xy}(T)}{\sigma_0})} \qquad (1),$$

where $M(T)$ and $M_0$ are the magnetization at temperature $T$ and saturation moment, respectively; $a_g$ and $b$ are nearly temperature independent constants, and $\sigma_0$ is a normalization factor which can be related to lattice parameter of the system [41]. Similarly, the spin wave stiffness $D_H(T)$ has a large temperature dependence that follows:

$$D_H(T) = \frac{\frac{a_D M(T)}{M_0}}{1 + b\left(\frac{M(T)}{M_0}\right)\left(\frac{\sigma_{xy}(T)}{\sigma_0}\right)} \qquad (2)$$

where $a_D$ is a near temperature independent constant. The same temperature dependence of the AHE (*b*) is used to describe the stiffness softening, consistent with previous analyses in [9, 41].

For $Co_3Sn_2S_2$, inelastic neutron scattering experiments confirm the strong interplay between spin dynamics and AHC [9, 10]. To test what happens in $Co_3Sn_{2-x}In_xS_2$, we measured the in-plane spin wave dispersion around (0,0,3) in $x = 0.12$ sample. Inelastic neutron scattering measurements were performed on single crystals with $x = 0.12$ using the IN-8 and Thales triple-axis spectrometers at the Institut Laue-Langevin (ILL) in Grenoble, France [42, 43]. We used doubly focusing pyrolytic graphite monochromator and analyzer with PG(0,0,2) reflection. Several scans were performed with a Cu(200)/PG(004) configuration. About 40 individual single crystals were co-aligned on an aluminum plate to form an assembly with a mosaic spread of 2 deg. The crystal assembly oriented in the [*H*, 0, *L*] horizontal scattering plane.

Figure 4a shows several constant-**Q** scans at (0,0,3), revealing clear spin gap at different temperatures. The spin gap was determined by fitting each curve to an ex-Gaussian function (detailed in the Methods Section). The resulting temperature dependence of the spin gap is shown in Fig. 4b where the blue dashed line represents the fit of eq. (1) with $a_g = 3.65$ meV and *b* = 0.60 compared to the black dashed line with $a_g = 2.36$ meV and *b* = 0.0 whose fitting is worse than with non-zero *b*. Therefore, like in the parent compound, the temperature dependence of the AHE and therefore non-zero *b* must be included to best fit the data. We compare the magnitude of *b* to that reported for the parent compound where *b*=0.39, indicating a strong interplay between the Weyl topology and spin dynamics with increased In doping [9].

Figure 4c shows the in-plane spin wave dispersions at 5 K and 200 K compared to the parent compound results of Ref. [9]. The temperature dependences of spin wave stiffness were obtained by fitting the spin wave dispersion with $E \approx E_g(T) + D_H(T)q^2$ where *E* is energy of spin waves, $E_g$ was fixed based on the ex-Gaussian fit results, and $q$ the momentum transfer away from the zone center. The resulting temperature dependence is shown in Figs. 4d. The black and blue dashed lines are fits of the $D_H(T)$ by eq. (2) with zero and finite $b$, respectively. Although finite $b$ fits the data slightly better, the differences with zero $b$ is not large.

**Discussion**

In a recent μSR study, the temperature and In-doping dependent magnetic phase diagram of $Co_3Sn_{2-x}In_xS_2$ was reported [29]. The FM ground state at $x = 0.0$ transitions into mixed FM+AFM state around 5 and 15% of In-doping, and then becomes FM+helimagnetic (HM) state for doping between 15 to 30%, and finally changes to HM state for $x \geq 0.30$. The outcome from our neutron diffraction results indicates that the phase diagram of $Co_3Sn_{2-x}In_xS_2$ is much simpler, showing only the FM+AFM phase with $\psi_2$ magnetic structure. The effect of In-doping is to transform $c$ axis FM order into in-plane AFM order. The fact that $x = 0.12$ sample has AFM order appearing well below the FM ordering temperature provided a natural interpretation of the temperature dependent AHE, suggesting the presence of THE from noncollinear spin structure contribution. These results, together with inelastic scattering measurements, clarified the phase diagram of the system, and provided the basis to understand the interplay between magnetism and topological magnetic properties.

In summary, we have used neutron diffraction to map out the In-doping evolution of the magnetic structures of $Co_3Sn_{2-x}In_xS_2$. We show that the parent compound is a pure ferromagnet with moment along the $c$ axis and $T_C = 177$ K. Around $x = 0.12$, the system is still a ferromagnet below 165 K, but an in-plane AFM order with $\psi_2$ magnetic structure sets in at $T_N = $ 50 K, forming a FM+AFM ($\psi_2 + \psi_3$) magnetic structure that induced a THE effect consistent with transport measurements. Finally, on moving to $x = 0.3$, AFM component increases at the expense of FM component, but the magnetic structure is still FM+AFM with $\psi_2 + \psi_3$ structure. Our inelastic neutron scattering experiments on $x = 0.12$ reveal that spin dynamics in the doped compound behave similarly as the undoped $Co_3Sn_2S_2$ where the magnon gap due to SOC and in-plane spin stiffness are renormalized by the same factor consistent with the AHE. The appearance of AFM order with noncollinear structure clearly affects the observed AHE. These results thus established the magnetic structures of the system from which a future microscopic theory for the AHE and THE can be established.

**Methods**

**Sample synthesis**

Single crystals of $Co_3Sn_{2-x}In_xS_2$ (0≤ $x$ ≤ 0.3) were grown by an Sn/In self flux method [21]. Initial compositions with the molar ratio Co:S:Sn/In = 8:6:86 were used. Co, Sn, S, and In powders were mixed together and placed in the bottom of an $Al_2O_3$ crucible. The top of the crucible was filled with quartz wool. The crucible was sealed in an evacuated quartz tube and placed in a box furnace. The mixture was heated to 400 ℃ over 2 hours, held for 2 hours, heated to 1050 ℃ over 6 hours, held for 6 hours, and them cooled slowly to 700 ℃ over 100 to 150 hours. The tube was removed from the furnace and spun in a centrifuge to remove flux. Plate-like crystals with hexagonal shape were obtained.

**Chemical composition**

The chemical composition of synthesized crystals was measured by X-ray diffraction (XRD) and Inductively Coupled Plasma Optical Emission Spectroscopy chemical analysis both at Rice University. Powder XRD measurements were performed on crushed single using a Rigaku Smartlab II X-ray Diffractometer to confirm the 322-phase. The In composition was determined using a Perkin Elmer Optima 8300 Inductively Coupled Plasma Optical Emission Spectrometer (ICP-OES) at Rice University. By comparing the ratio of Co:In in dissolved samples, an average doping level of $x = 0.12$. was found over four samples from four different batches with $x_{nom} = 0.15$. Similar analysis found $x = 0.3$ for batches with $x_{nom} = 0.4$.

**Magnetization measurements**

The field-cooled (FC) temperature dependent magnetization data was fit to a power law function to extract the value of $T_C$ excluding the low temperature data in $x = 0.12$ data. Then including all data, an additional fit to a polynomial function was used to fully capture the measured temperature dependence and compared to the temperature dependence of energy gap and in-plane dispersion (Figures 4b and 4d in the main text). An example of temperature dependent magnetization is shown in Supplementary Figure 1 for $x = 0.3$ sample. Here we compare zero-field cooled (ZFC) and FC data and clearly see the domain size transition $T_D$ near 120 K.

Measurements were done to confirm the In-doping level by comparing to previous studies of the doping dependence of $T_C$ by [19]. Based on those previous results, $T_C$ follows a doping dependence approximately described by the function $T_C = -59.51 * x^2 - 134.1 * x + 182.1$ as shown in Supplementary Figure 2. Using this function to compare $T_C$ in our samples indicated doping levels of $x = 0.12$ and $x = 0.3$. These levels are consistent with our ICP-OES results. Additionally, the magnitude of the anomalous Hall conductivity observed in [19] are consistent with our results.

**Transport measurements**

Transport measurements were performed on samples polished and cut to a bar shape. Silver paste and gold wires were used to make resistivity measurements in a Hall bar geometry as shown in Supplementary Figure 3a-b. Temperature dependent resistivity measurements are shown in Supplementary Figure 4 as a function of temperature. The temperature dependences of anomalous Hall resistivity and conductivity were fit to a polynomial function and compared to the temperature dependence of energy gap, in-plane dispersion, and anomalous Hall resistivity. Temperature dependence of the percent deviation between the fit and measured anomalous Hall resistivity is shown in Supplementary Figure 5.

**Elastic neutron scattering**

Our elastic neutron scattering experiments were carried out at CORELLI at the Spallation Neutron Source (SNS) and WAND$^2$ at the High Flux Isotope Reactor (HFIR) at Oak Ridge National Laboratory (ORNL), Oak Ridge, TN, USA [25, 41]. For each experiment and sample doping level, a single crystal of 10-50 mg was mounted in the [$H$,0,$L$] scattering plane at CORELLI and [$H$,-$H$,0] scattering plane at WAND$^2$. The crystals were put in a cryostat and rotated 360 degrees to collect large reciprocal space maps as a function of temperature. The Mantid package was used for data reduction and analysis. Jana2006 software was used for structure refinement. Nuclear refinements were done using peak intensities from the 165 K, $x =$ 0.12 and 160 K, $x = 0.3$. A total of 304 and 150 reflections, respectively, were used in the refinement and resulted in good fits with $R$ values of 3.02 and 4.55. These nuclear refinements were used as inputs for the magnetic refinements at 6 and 60 K for $x = 0.12$ and 6 K for $x = 0.3$. For $x = 0.12$ 6 K, 408 reflections were used and resulted in an R value of 3.80. For $x = 0.12$ 60 K, 369 reflections were used and resulted in an $R$ value of 3.63. For $x = 0.3$ 6 K, 168 reflections were used and resulted in an $R$ value of 4.04.

**Representation analysis**

Based on the magnetic Co sublattice and the **k**=0 propagation vector, there are nine possible magnetic configurations as shown in Table 1. The nine basis vectors, $\psi_{1-9}$, are categorized into three irreducible representations (irreps) labeled by $\Gamma^+_{1-3}$. The moment direction are relative magnitude are **M**(x,y,z).

**Inelastic neutron scattering**

Inelastic neutron scattering measurements were performed on single crystals with $x = 0.12$ using the IN-8 and Thales spectrometers at the Institut Laue-Langevin (ILL) in Grenoble, France. We used doubly focusing pyrolytic graphite monochromator and analyzer with PG(0,0,2) reflection. Several scans have been performed with a Cu(200)/PG(004) configuration. For these experiments, ~1 g of $x = 0.12$ crystals were coaligned in the [$H$,0,$L$] scattering plane and fixed to an Al plate as shown in Supplementary Figure 6.

The temperature dependence of the energy gap was measured by performing constant-**Q** cuts from 0 to 5 meV. To extract the energy gap, each leading-edge peak was fit to an exponentially modified Gaussian distribution, a convolution of a Gaussian and exponential distribution, where the exponential distribution accounts for the asymmetric tail that results from finite resolution. The exponentially modified Gaussian distribution can be written as, $f(x) = \frac{\lambda}{2} e^{\frac{\lambda}{2}(2\mu+\lambda\sigma^2-2x)} erfc \frac{\mu+\lambda\sigma^2-x}{\sqrt{2}\sigma}$,, with parameters $(\lambda, \sigma, \mu)$ where $\lambda$ is a constant, $\sigma$ is the standard deviation of the Gaussian, and $\mu$ is the mean of the Gaussian. For each temperature [See Supplementary Figure 7a], this function was fit to the data [See Supplementary Figure 7b] and the peak position was extracted as the energy gap ($E_g$) as summarized in Table 2.

Constant energy scans were performed along [0,0,$L$] and [$H$,0,3] as a function of temperature. Several scans are shown in Supplementary Figure 8. For each scan, data were fit to a double Gaussian function including two slightly asymmetric peaks as

$$I = I_A \left[ e^{-(H-\delta)^2/2\sigma^2} \right] + I_B \left[ e^{-(H+\delta)^2/2\sigma^2} \right]$$

for $H$-scans and

$$I = I_A \left[ e^{-(L-3-\delta)^2/2\sigma^2} \right] + I_B \left[ e^{-(L-3+\delta)^2/2\sigma^2} \right]$$

for $L$-scans after subtracting a linear background. At each temperature, energy versus $\pm\delta$ was fit to the function $E = E_g + Dq^2$ using the measured values of $E_g$ to determine the spin stiffness $D$.

**Data availability**

The data supporting the findings of this study are available within the paper and in the Supplementary Information. The raw data are available from the corresponding authors upon reasonable request. The raw data obtained on IN8 and Thales at ILL are available at https://doi.ill.fr/10.5291/ILL-DATA.4-01-1630. The raw data from CORELLI and the WAND$^2$ diffractometers are available upon request.

**Acknowledgements**

We wish to thank Ursula Bengaard Hansen and Paul Steffens for help and assistance for the Thales experiment. The neutron scattering and single crystal synthesis work at Rice is supported by US NSF-DMR-2100741 and by the Robert A. Welch Foundation under Grant No. C-1839 (P.D.). The transport measurement and single crystal synthesis work at UW were supported by the Air Force Office of Scientific Research (AFOSR) under Award No. FA2386-21-1-4060 and the David Lucile Packard Foundation (J.H.C). A portion of this research used resources at the Spallation Neutron Source and the High Flux Isotope Reactor, a DOE Office of Science User Facility operated by ORNL.

Competing interests

The authors declare no competing interests.

Author contributions

P.D. and J.-H.C. conceived and managed the project. The single-crystal $Co_3Sn_{2-x}In_xS_2$ samples were grown by K.J.N. and B.G. Neutron diffraction experiments at CORELLI and the WAND$^2$ diffractometers were performed by K.J.N., F.Y. and K.M.T.. Neutron refinements were carried out by K.J.N. in discussion with F.Y. Inelastic neutron scattering experiments at IN-8 and Thales spectrometers were performed by P.B. and A.I. in discussion with P.D. and K.J.N.. Magnetic susceptibility measurements were performed by K.J.N.. Transport measurements were carried out by P.M. and J-H.C. The paper is written by K.J.N., P.D., J.-H.C., and all co-authors made comments on the paper.

**References**

1. Hasan, M. Z. & Kane, C. L. *Colloquium*: topological insulators. *Rev. Mod. Phys.* **82**, 3045 (2010).

2. Qi, X. -L. & Zhang, S. -C.Topological insulators and superconductors. *Rev. Mod. Phys.* **83**, 1057 (2011).

3. Wang, Q. H. et al. The magnetic genome of two-dimensional van der Waals materials", *ACS Nano* **16**, 5, 6960-7079 (2022).

4. Lv, B. Q., Qian, T. & Ding, H. Experimental perspective on three-dimensional topological semimetals. *Rev. Mod. Phys.* **93**, 025002 (2021).

5. Liu, E. et al. Giant anomalous Hall effect in a ferromagnetic kagome-lattice semimetal. *Nat. Phys.* **14**, 1125-1131 (2018).

6. Wang, Q. et al. Large intrinsic anomalous Hall effect in half-metallic ferromagnet $Co_3Sn_2S_2$ with magnetic Weyl fermions. *Nat. Commun.* **9**, 3681 (2018).

7. Ghimire, M. P. et al. Creating Weyl nodes and controlling their energy by magnetization rotation. *Phys. Rev. Research* **1**, 032044 (2019).

8. Yang, R. et al. Magnetization-induced band shift in ferromagnetic Weyl semimetal $Co_3Sn_2S_2$. *Phys. Rev. Lett.* **124**, 077403 (2020).

9. Liu, C. et al. Spin excitations and spin wave gap in the ferromagnetic Weyl semimetal $Co_3Sn_2S_2$. *Sci. China: Phys. Mech. Astron.* **64**, 217062 (2021).

10. Zhang, Q. et al. Unusual exchange couplings and intermediate temperature Weyl state in $Co_3Sn_2S_2$. *Phys. Rev. Lett.* **127**, 117201 (2021).

11. Guguchia, Z. et al. Tunable anomalous Hall conductivity through volume-wise magnetic competition in a topological kagome magnet. *Nat. Commun.* **11**, 559 (2020).
18

42. Boehm, M. et al. ThALES—Three axis low energy spectroscopy for highly correlated electron systems. *Neutron News* **26**, 18-21 (2015).

43. Hiess, A. et al. ILL's renewed thermal three-axis spectrometer IN8: A review of its first three years on duty. *Phys. B: Condens. Matter* **385-386**, 1077-1079 (2006).
22

Figures

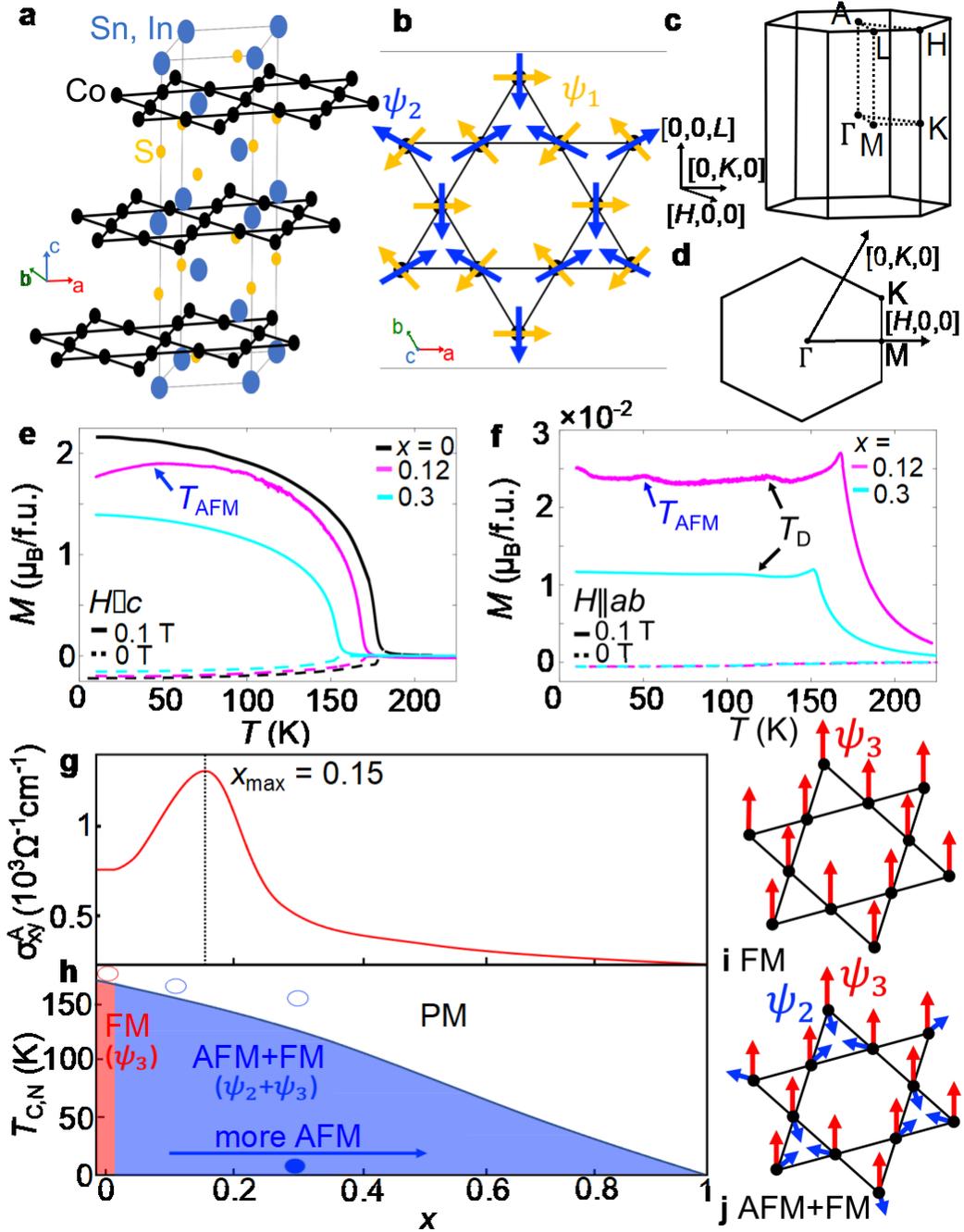

**Fig. 1: Crystal and magnetic properties of Co$_3$Sn$_{2-x}$In$_x$S$_2$**

**a** Crystal structure of Co$_3$Sn$_2$S$_2$ from **a** three-dimensional view. **b** Top view with the in-plane symmetry-allowed spin configurations $\psi_1$ and $\psi_2$. **c-d** Brillouin zone boundaries and high-symmetry points in a **c** three-dimensional and **d** two-dimensional view. **e-f e** Out-of-plane and **f** in-plane magnetization as a function of doping in zero and low field. (g) Indium doping dependence of the anomalous Hall conductivity based on [19]. **h** Temperature-doping phase diagram. Dots indicate temperature and doping of neutron experiments. **i-j** In-plane and out-of-plane components of the refined magnetic structures at $x =$ **i** 0.0 and **j** 0.12 and 0.3.

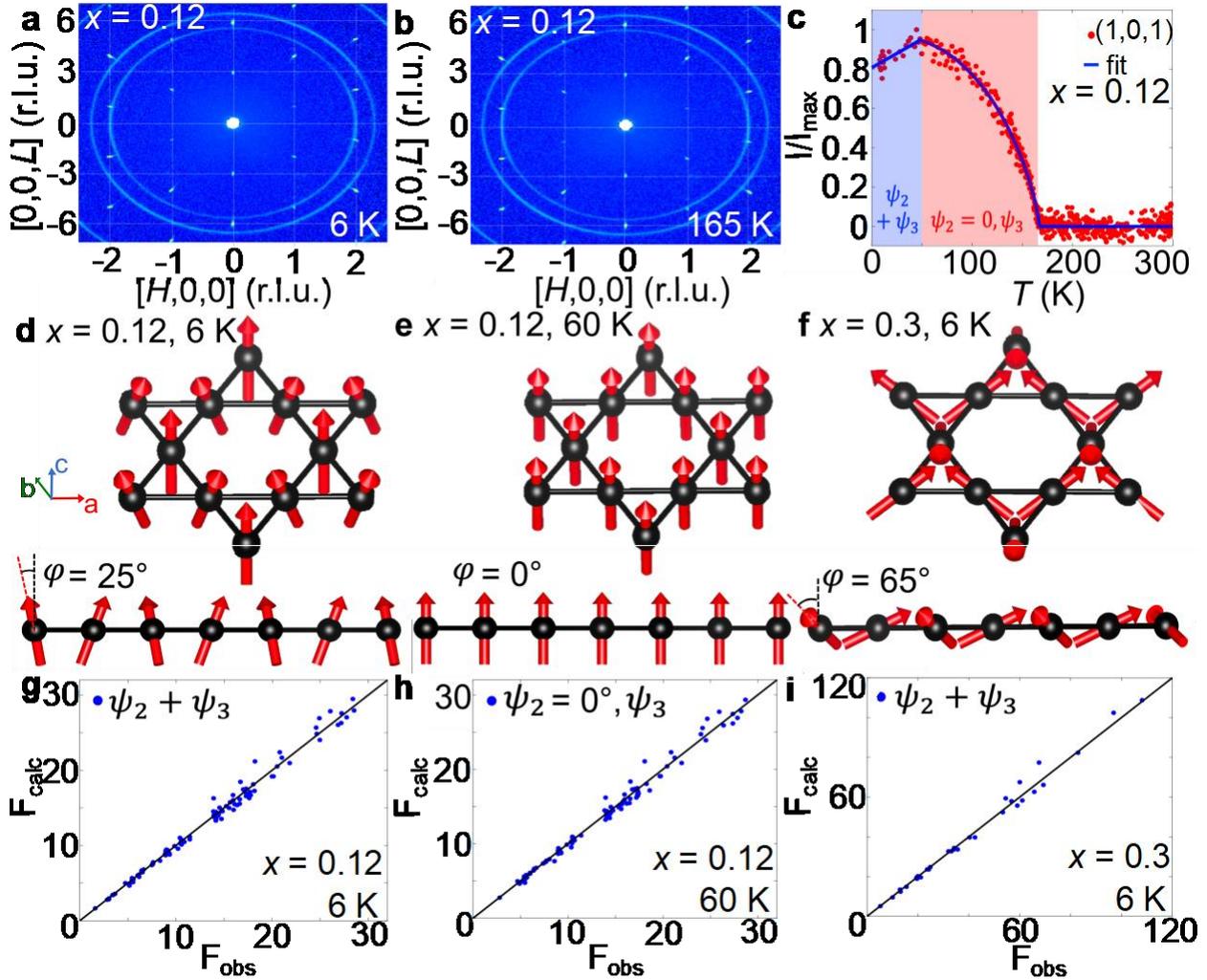

**Fig. 2: Neutron diffraction results and magnetic structures of $Co_3Sn_{2-x}In_xS_2$**

**a-b** Maps of magnetic peaks in reciprocal space from neutron diffraction of $x = 0.12$ sample at **a** 6 K and **b** 165 K. **c** Temperature dependence of the (1,0,1) magnetic Bragg peak intensity. **d-f** Refined magnetic structures in $x = 0.12$ crystals at **d** 6 K and **e** 60 K and **f** $x = 0.3$ at 6K. **g-i** Comparison of measured ($F_{obs}$) and calculated ($F_{calc}$) integrated intensities from the refinement in
24

the $\psi_2 + \psi_3$ magnetically ordered phases shown in **d-f** in $x = 0.12$ crystals at **g** $T = 6$ K and **h** 60 K and **i** $x = 0.3$ at 6 K. The overall $R$ values = 3.79, 3.63, and 4.04 for $x = 0.12$ at **g** $T = 6$ K, **h** 60 K, and **i** $x = 0.3$ at 6 K, respectively. The overall $R$ values best fit the data for the $\psi_2 + \psi_3$ structure compared to the $\psi_1 + \psi_3$ structure where overall $R = 4.70$, 4.58, and 5.50 for $x = 0.12$ at **g** T = 6 K, **h** 60 K, and **i** $x = 0.3$ at 6 K, respectively.

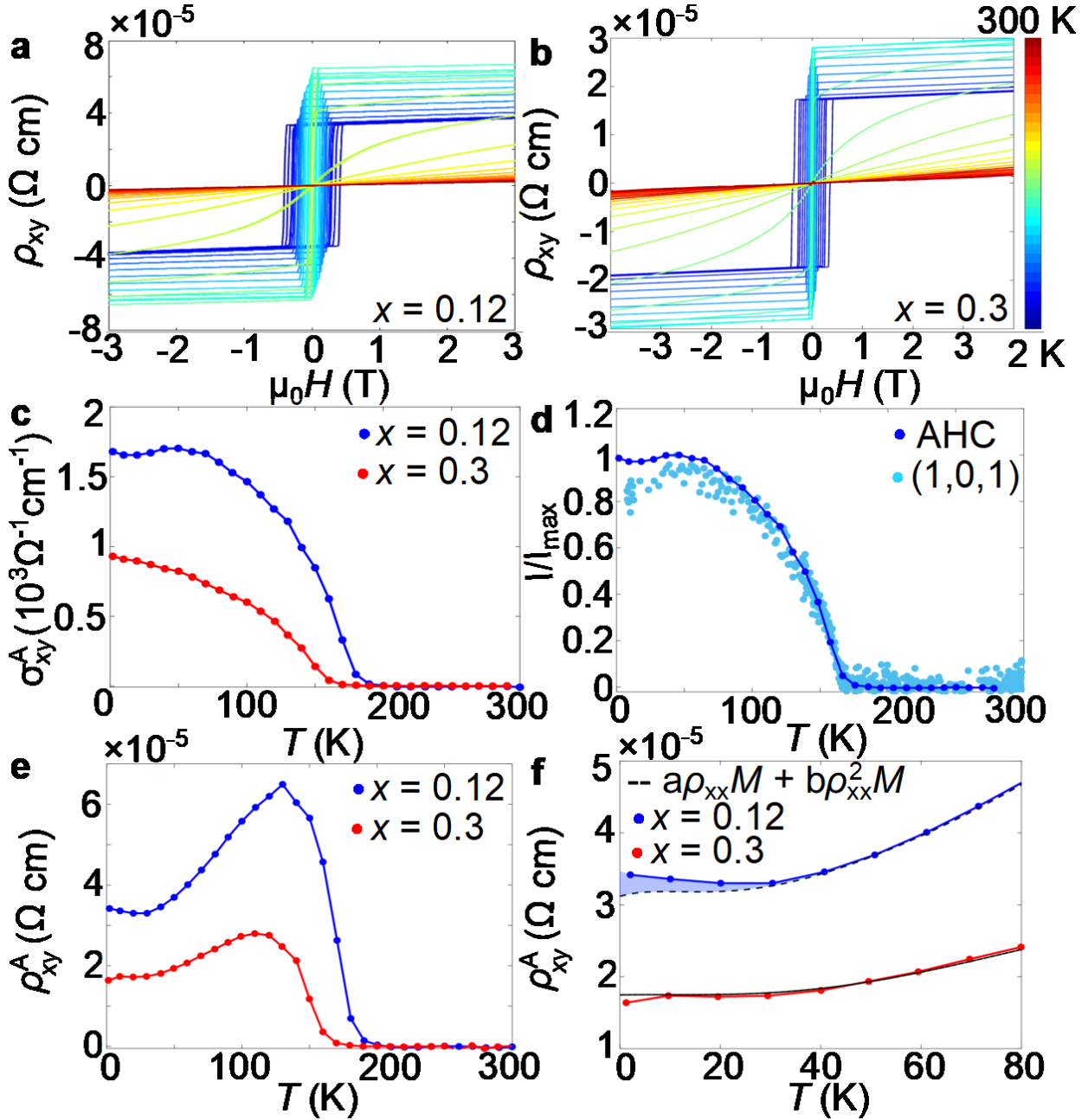

**Fig. 3: Transport measurements of $Co_3Sn_{2-x}In_xS_2$**

**a-b** Field dependence of $\rho_{xy}$ for **a** $x = 0.12$ and **b** $x = 0.3$. **c** Temperature dependence of the anomalous Hall conductivity. **d** Anomalous Hall conductivity compared to order parameter of (1,0,1) magnetic peak. **e** Temperature dependence of the anomalous Hall resistivity. **f** Anomalous Hall resistivity compared fit to the expected contributions from intrinsic and extrinsic scattering mechanisms. Deviations at low temperature in the $x = 0.12$ sample (shaded in blue) indicate a possible topological Hall contribution due to non-zero spin chirality.

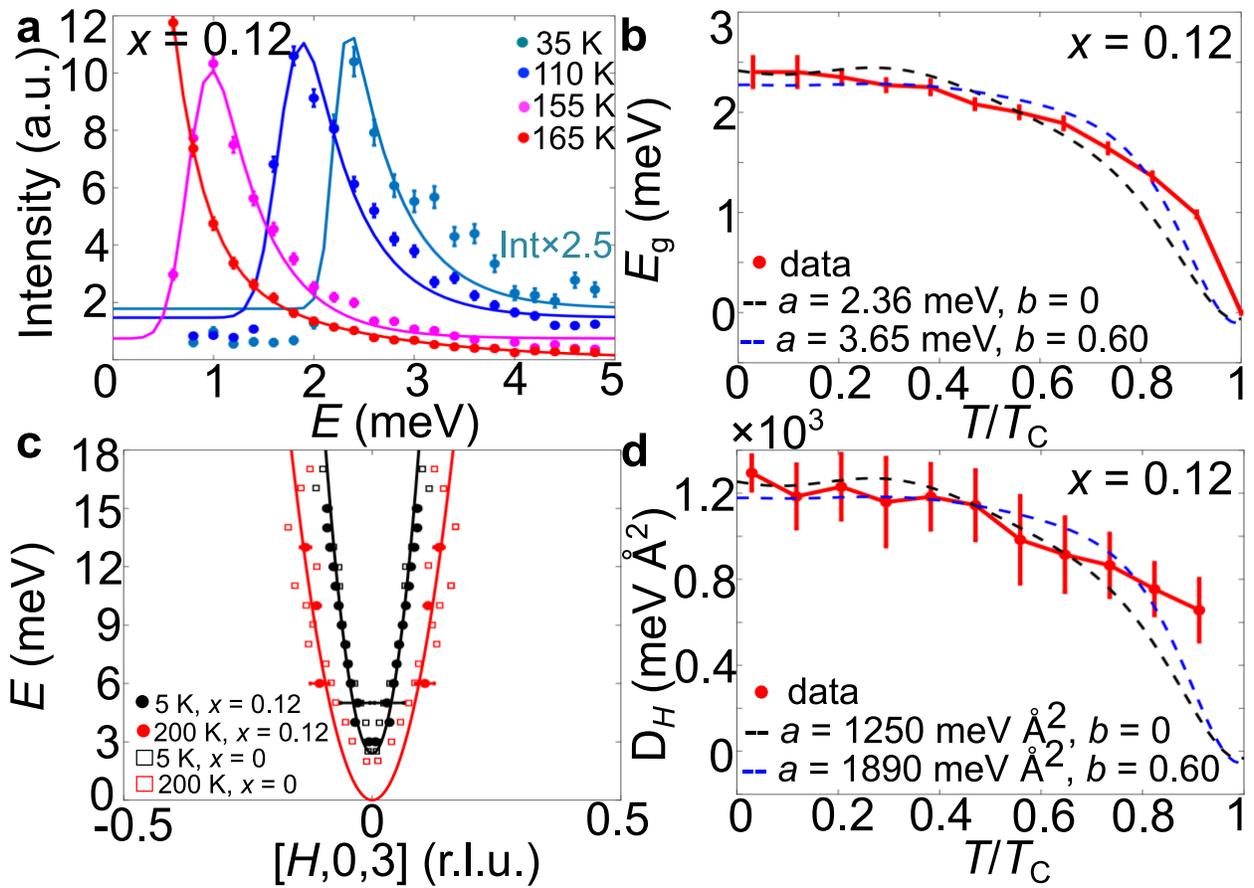

**Fig. 4: Inelastic neutron scattering results of $Co_3Sn_{2-x}In_xS_2$ ($x = 0.12$)**

**a** Select energy dependence of the spin excitations at $T = 35, 110, 155$, and $165$ K (see supplementary information for fit details). **b** Temperature dependence of the spin wave gap and fitting results with different parameters with $T_C = 165$ K. **c** Dispersion curve of the spin wave excitations along [$H$,0,3] at 5 and 200 K fit to a $q^2$-dependence and compared to $x = 0$ data from [9] as shown in square points. After data analysis, [9] claims a gapless spectrum at 200 K. **d** Temperature dependence of the in-plane spin-wave stiffness $D_H$ and fitting results with different parameters and $T_C$ = 165 K. $D_H$ was determined using fixed $E_g$ determined from energy scans (Fig. 4a) and then fitting the dispersions along $H$ (Fig. 4c). The vertical error bars in **a** are statistical errors of 1 standard deviation. The horizontal and vertical error bars in **b**-**d** represent the estimated uncertainty obtained in fitting procedures.

**Tables**

**Table 1 The irreducible representations allowed by symmetry and their corresponding basis vectors and moment components**

| No. | Irrep | Basis Vectors | M (x,y,z) | | |
|---|---|---|---|---|---|
| | | | (x,y,z) = (1/2,1/2,1/2) | (1/2,0,1/2) | (0,1/2,1/2) |
| 1 | $\Gamma_1^+$ | $\psi_1$ | $\mathbf{M} = \begin{pmatrix} a \\ a \\ 0 \end{pmatrix}$ | $\begin{pmatrix} -a \\ 0 \\ 0 \end{pmatrix}$ | $\begin{pmatrix} 0 \\ -a \\ 0 \end{pmatrix}$ |
| 2 | $\Gamma_2^+$ | $\psi_2$ | $\begin{pmatrix} a \\ -a \\ 0 \end{pmatrix}$ | $\begin{pmatrix} a \\ 2a \\ 0 \end{pmatrix}$ | $\begin{pmatrix} -2a \\ -a \\ 0 \end{pmatrix}$ |
| 3 | | $\psi_3$ | $\begin{pmatrix} 0 \\ 0 \\ 1 \end{pmatrix}$ | $\begin{pmatrix} 0 \\ 0 \\ 1 \end{pmatrix}$ | $\begin{pmatrix} 0 \\ 0 \\ 1 \end{pmatrix}$ |
| 4 | $\Gamma_3^+$ | $\psi_4$ | $\begin{pmatrix} 1 \\ 0 \\ 0 \end{pmatrix}$ | $\begin{pmatrix} 0 \\ b^* \\ 0 \end{pmatrix}$ | $\begin{pmatrix} -b \\ -b \\ 0 \end{pmatrix}$ |
| 5 | | $\psi_5$ | $\begin{pmatrix} 1 \\ 0 \\ 0 \end{pmatrix}$ | $\begin{pmatrix} 0 \\ b \\ 0 \end{pmatrix}$ | $\begin{pmatrix} -b^* \\ -b^* \\ 0 \end{pmatrix}$ |
| 6 | | $\psi_6$ | $\begin{pmatrix} 0 \\ 1 \\ 0 \end{pmatrix}$ | $\begin{pmatrix} -b^* \\ -b^* \\ 0 \end{pmatrix}$ | $\begin{pmatrix} b \\ 0 \\ 0 \end{pmatrix}$ |
| 7 | | $\psi_7$ | $\begin{pmatrix} 0 \\ 1 \\ 0 \end{pmatrix}$ | $\begin{pmatrix} -b \\ -b \\ 0 \end{pmatrix}$ | $\begin{pmatrix} b^* \\ 0 \\ 0 \end{pmatrix}$ |
| 8 | | $\psi_8$ | $\begin{pmatrix} 0 \\ 0 \\ 1 \end{pmatrix}$ | $\begin{pmatrix} 0 \\ 0 \\ b^* \end{pmatrix}$ | $\begin{pmatrix} 0 \\ 0 \\ b \end{pmatrix}$ |

| 9 | | $\psi_9$ | $\begin{pmatrix} 0 \\ 0 \\ -1 \end{pmatrix}$ | $\begin{pmatrix} 0 \\ 0 \\ -b \end{pmatrix}$ | $\begin{pmatrix} 0 \\ 0 \\ -b^* \end{pmatrix}$ |

**Table 2 Experimental gap energies determined by ex-gaussian fittings for 35≤ $T$ ≤155 K**

| T (K) | $E_g$ (meV) | $D_H$ (meV Å$^2$) |
| --- | --- | --- |
| 5 | 2.40±0.15 | 1780±110 |
| 20 | 2.40±0.15 | 1630±200 |
| 35 | 2.35±0.07 | 1290±210 |
| 50 | 2.27±0.06 | 1590±280 |
| 65 | 2.25±0.07 | 1630±210 |
| 80 | 2.08±0.05 | 1570±220 |
| 95 | 2.00±0.06 | 1350±280 |
| 110 | 1.89±0.06 | 1260±240 |
| 125 | 1.64±0.05 | 1190±200 |
| 140 | 1.36±0.04 | 1040±170 |
| 155 | 0.98±0.03 | 900±200 |